# Achieving low contact resistance through copper-intercalated bilayer MoS$_2$

Huan Wang[1,2], Xiaojie Liu[1*], Hui Wang[1*], Yin Wang and Haitao Yin[1*]

1. Key Laboratory for Photonic and Electronic Bandgap Materials of Ministry of Education, School of Physics and Electronic Engineerin, Harbin Normal University, Harbin, 150025, China

2. College of Science, Qiqihar University, Qiqihar, 161006, China

3. Department of Physics and International Centre for Quantum and Molecular Structures, Shanghai University, Shanghai, 200444, China

**ABSTRACT**

The high contact resistance between MoS$_2$ and metals hinders its potential as an ideal solution for overcoming the short channel effect in silicon-based FETs at sub-3nm scales. We theoretically designed a MoS$_2$-based transistor, featuring bilayer MoS$_2$ connected to Cu-intercalated bilayer MoS$_2$ electrodes. At 0.6 V, contact resistance is 16.7 Ω·$\mu$m (zigzag) and 30.0 Ω·$\mu$m (armchair), nearing or even surpassing the 30 Ω·$\mu$m quantum limit for single-layer materials. This low resistance is attributed to the elimination of the tunneling barrier and the creation of ohmic contacts. Additionally, the small contact potential difference enables lower operating voltages. Our intercalation design offers a novel approach to achieving low contact resistance in 2D electronic devices.

## I. INTRODUCTION

In the process of semiconductor applications, it is inevitable to involve the contact between metal electrodes and semiconductors. Through electrical contact, charge carriers are injected from the metal into the semiconductor material. This contact significantly influences the performance properties of the device. Achieving good electrical contact is a prerequisite for two-dimensional (2D) materials to unlock their full potential.[1] The quality of the electrical contact can be evaluated by measuring the contact resistance ($R_C$). Contact resistance arises because carrier injection into 2D semiconductor materials through electrical contact is restricted. The Schottky barrier and tunneling barrier, which influence carrier injection, are crucial for achieving low contact

---

[*]Corresponding author: wlyht@126.com (Haitao Yin), redlxj@126.com (Xiaojie Liu),wh@fysik.cn (Hui Wang)

resistance.

In the ideal metal-semiconductor junction, according to the Schottky-Mott law,[2] the height of the Schottky barrier can be regulated by selecting an appropriate work function of the metal.[3–5] Due to factors such as Fermi pinning[6] and chemical bonding between metals and semiconductors, the height of the Schottky barrier no longer follows the Schottky-Mott law.[7] The absence of suspensions in 2D materials avoids chemical disorder and defect/metal-induced gap states, greatly suppresses interfacial dipoles, and overcomes Fermi pinning effects.[8–15] However, the weak hybrid coupling of the wave function between metal and 2D semiconductor interface makes it difficult to form a strong interfacial chemical bond with metal. The additional tunnel barrier introduced by the Van der Waals gap reduces charge injection and increases contact resistance. In addition to the top contact, the 2D metal and semiconductor contact modes also include edge contact. The edge contact creates a stronger hybrid at the metal interface than the top contact, eliminating the additional tunnel barrier. But it is challenging to fabricate pure edge contacts with standard lithography techniques in 2D materials.

As a suitable channel material, $MoS_2$ is naturally abundant and can be synthesized on various insulating substrates using chemical vapor deposition (CVD) technology.[16–18] In terms of electrical performance, $MoS_2$ has a suitable band gap, high switching ratio and carrier mobility. Recent advancements incorporating novel materials and contact engineering techniques have enabled the reduction of $R_C$ below 1 k$\Omega$·$\mu$m.[19–22] The 2H-$MoS_2$ semiconductor is transformed into 1T-$MoS_2$ by phase engineering technology. The $R_C$ of the 1T-$MoS_2$/2H-$MoS_2$ junction is 200 $\Omega$·$\mu$m.[23] Furthermore, it has been demonstrated that semimetals such as Bi and Sb, capable of effectively suppressing metal-induced gap states (MIGS) owing to their low density of states (DOS) at the Fermi level, facilitate the formation of high-quality n-type ohmic contacts with various transition metal dichalcogenides (TMDs). Notably, the $R_C$ values achieved are as low as 123 $\Omega$·$\mu$m for the Bi/$MoS_2$ interface and 42 $\Omega$·$\mu$m for the Sb/$MoS_2$ interface.[8,24] Utilizing semimetal as the electrode, while effectively decreasing the contact resistance, unfortunately introduces a drawback: a higher operating voltage is necessary to achieve a larger on-state current.

The intercalation method has surfaced as a promising approach to tackle the challenge of elevated contact resistance between TMDs and conventional metals. In the case of Van der Waals materials, such as $MoS_2$, there exists considerable interlayer spacing, and harnessing this space can effectively modulate the material's characteristics. A common approach to achieve this is through intercalation technology, which entails inserting atoms into the interlayer spaces of the host material. The charge transfer facilitated by the inserted atoms or the heightened spin-orbit coupling resulting from the incorporation of heavy atoms can significantly boost superconductivity,[25,26] thermoelectricity,[27] spin polarization,[28,29] and optoelectronic properties.[30] For instance, the intercalation of Li and Fe atoms into bilayer $2H$-$TaS_2$ leads to the induction of superconductivity and ferromagnetism,[25,29] respectively.

Furthermore, the intercalation of Co/Cu atoms into bilayer $SnS_2$ (Co/Cu-$SnS_2$) can locally convert this n-type semiconductor into a p-type semiconductor or even a metal.[31] Additionally, the Co-$SnS_2$/$SnS_2$ junction attains a notably low contact resistance.[32] For $MoS_2$ materials, which garner significant research attention, it is imperative to identify metal-like electrode materials through the application of intercalation techniques. By constructing seamless metal/$MoS_2$ junctions, Fermi level pinning can be eliminated, paving the way for achieving ultra-low contact resistance, a critical factor in enhancing the performance of $MoS_2$-based field-effect transistors.

In this work, we unveil our prediction of a metallic Cu-intercalated bilayer $MoS_2$ (Cu-$MoS_2$) structure and introduce a novel design for a field-effect transistor (FET) that employs this structure as the electrode, with bilayer $MoS_2$ (BL-$MoS_2$) acting as the channel. We have attained contact resistances along both the armchair (AC) and zigzag (ZZ) directions that are either near or below the quantum limit value of 30 $\Omega \cdot \mu m$ for single-layer materials. The gate voltage ($V_{gs}$) modulation effect is profound, with the On/Off state currents satisfying the high-performance (HP) standard of 900 $\mu A/\mu m$ and 0.1 $\mu A/\mu m$, as stipulated in the International Technology Roadmap for Semiconductors (ITRS) for the year 2028.

## II. MODEL AND METHODS

Figure 1(a) and 1(c) illustrate the crystal structures of BL-$MoS_2$ and Cu-$MoS_2$, respectively. The unit cell lattice is represented by a tetragonal prism with parameters a=b=3.16 Å and c=23 Å, including a vacuum layer of 13 Å thickness. BL-$MoS_2$ is stacked in A and A' modes with a layer interval of 7.05 Å. The Cu atoms are positioned directly below the topmost S atom and above the bottommost Mo atom, maintaining respective distances of 2.20 Å from the upper $MoS_2$ layer and 1.61 Å from the lower $MoS_2$ layer, respectively. These distances were optimized using VASP software[33] at the generalized gradient approximation (GGA) level after considering the Van der

Waals correction with DFT-D3 method of Grimme with zero-damping function.[34] The Brillouin zone was sampled with a k-point of 4 × 1 × 4. The plane wave cut-off energy was set to 400 eV, and the convergence accuracy of force and energy were 0.02 eV/Å and 1×10$^{-8}$ eV/atom, respectively. Figures 1(b) and 1(d) show the band structures of BL-MoS$_2$ and Cu-MoS$_2$, respectively, as calculated by DFT. As shown in Figure 1(b), BL-MoS$_2$ is a direct bandgap semiconductor, which is consistent with previous experimental results.[35] For Cu-MoS$_2$, the band structure changes significantly due to the coupling between the Cu atom and MoS$_2$. As shown in Figure 1(d), the Fermi level intersects the energy bands, resulting in partial occupancy of the energy levels in the conduction band by electrons, while other levels remain unoccupied, indicating conducting behavior.

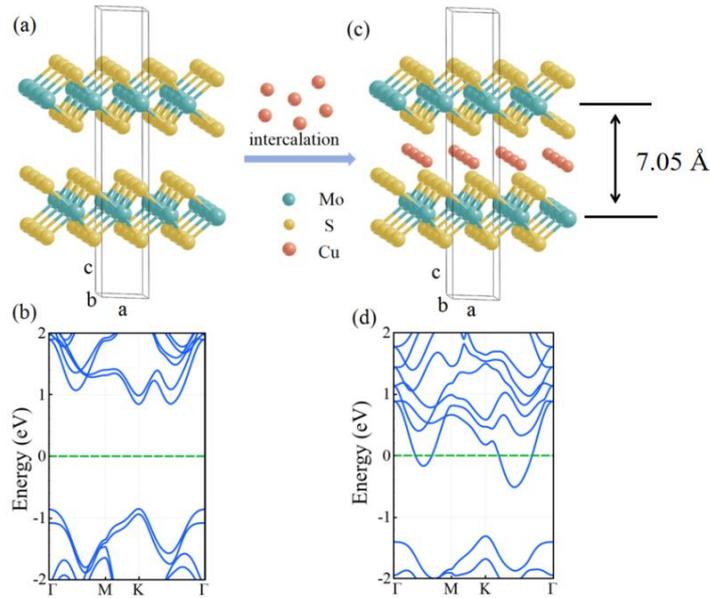

Figure 1. Geometric and band structures of bilayer MoS$_2$: Geometric structures of (a) bilayer MoS$_2$ stacked in A and A' configuration and (b) bilayer MoS$_2$ with Cu intercalation. Band structures of (c) bilayer MoS$_2$ unit cell and (d) bilayer MoS$_2$ unit cell with Cu intercalation.

The MoS$_2$-based FETs we constructed are shown in Figure 2(a) and 2(b). In these structures, the metallic Cu-MoS$_2$ serves as the electrode, with MoS$_2$ as the central region. The lengths of MoS$_2$ along the AC direction and the ZZ direction are 21 Å and 22 Å, respectively. The central region is covered with a 5 Å-thick HfO$_2$ dielectric layer. The atoms at the interface were locally optimized using the VASP software. The source and drain electrodes extend to the positive and negative infinity of the *z* axis, respectively, where a bias voltage is applied and current is collected. Both the bottom and top gates are positioned near the central scattering region. The transport properties of the device were calculated by the density functional theory (DFT) and the nonequilibrium Green's function (NEGF) method, which is implemented in NANODCAL software.[36] In the calculation, GGA was used to deal with the exchange correlation potential,[37] double zeta polarized atomic orbital group was used to expand wave function, and standard nonlocal mode conservation pseudopotential was used to replace the interaction between the nucleus and inner electron and valence electron.[38] In the process of self-consistent calculation and post-analysis, k-point sampling was 15×1×1.

Under the linear response approximation, the current flowing through the Cu-MoS$_2$/BL-MoS$_2$/Cu-MoS$_2$ device can be calculated using the Landauer-Büttiker formula,[39]

$$I(V_{ds}) = -\frac{2e}{h}\int_{\mu_L}^{\mu_R} T(E, V_{ds})\left[f_L(E, \mu_L) - f_R(E, \mu_R)\right]dE \tag{1}$$

where $f_L(f_R)$ represents the Fermi-Dirac distribution function and $\mu_L(\mu_R)$ denotes the electrochemical potential of the left (right) electrode. The relationship between $\mu_L(\mu_R)$ and the bias voltage applied between the source and drain electrodes, $V_{ds}$, satisfies $|\mu_L - \mu_R| = eV_{ds}$. The transmission coefficient $T(E, V_{ds})$ is defined as

$$T = \text{Tr}[\Gamma_L G^r \Gamma_R G^a] \tag{2}$$

where the line-width function $\Gamma_{L/R}$ describes the coupling between the electrode and the central scattering region. $G^r$ ($G^a$) is the retarded (advanced) Green's function.

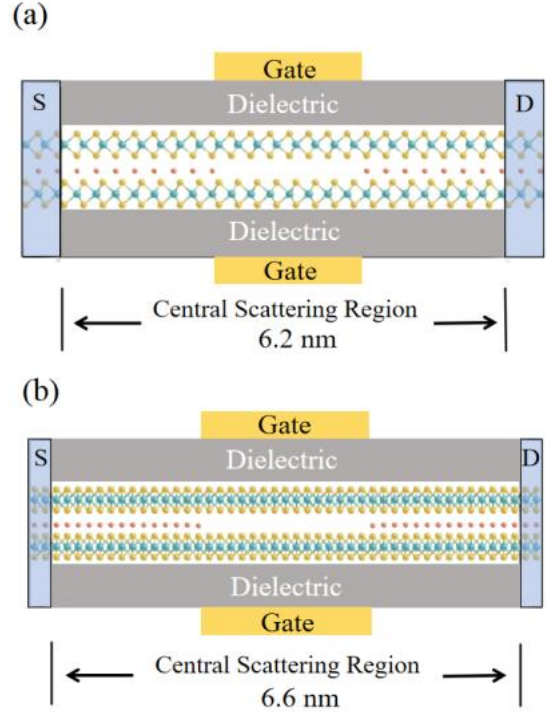

Figure 2 Schematic structures of the Cu-MoS$_2$/BL-MoS$_2$/Cu-MoS$_2$ devices along the AC direction (a) and the ZZ direction (b). S (D) denotes the source (srain) electrodes.

## III. RESULTS AND DISCUSSION

The obtained *I-V* curves of the Cu-MoS$_2$/BL-MoS$_2$/Cu-MoS$_2$ devices under different source-drain voltages $V_{ds}$ in both the AC and ZZ directions are shown in Figure 3(a). The currents increase nonlinearly with the increase of $V_{ds}$. According to the *I-V* curves, the contact resistance ($R_C$) can be calculated by the formula $R=V/2I$. Since the channel length of the device is shorter than the mean free path of the electrons, the electrons will not experience scattering during the transport process, so the resistance of the system is primarily due to the contact interface between metal and semiconductor. Figure 3(b) illustrates the $R_C$ observed at various $V_{ds}$ for both the AC and ZZ directions. As the $V_{ds}$ increases, the $R_C$ decreases accordingly. Specifically, at $V_{ds}$ = 0.6 V, the $R_C$ drops to 16.7 Ω·$\mu$m in the ZZ direction and 30 Ω·$\mu$m in the AC direction. This is significantly lower than the contact resistance of monolayer MoS$_2$ by in situ Fe doping device obtained by chemical vapor deposition technology, and it is reported to be 678 Ω·$\mu$m,[7] and the contact resistance of the In/Sn alloy in contact with the upper surface of monolayer MoS$_2$ is 190 Ω·$\mu$m.[21] The contact resistance we obtained is also smaller than the one obtained when using semimetal as the electrode.[8,24] These comparisons show that the proposed device designed using the intercalation method has low contact resistance even at small bias voltages. Moreover, the contact resistance in the ZZ (AC) direction is below (close to) the theoretical limit of 30 Ω·$\mu$m for single-layer materials.

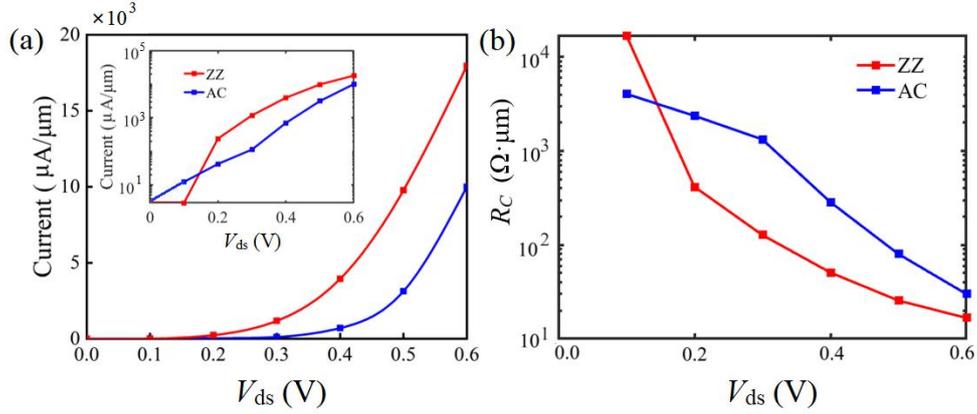

Figure 3 (a) I-V curves, and (b) contact resistance ($R_C$) versus bias voltage. Inset: I-V curves on log y scale.

In order to investigate the reason for low contact resistance, the projected density of states (PDOS) at equilibrium were calculated. The left panels of Figure 4(a) and 4(b) show the PDOS on the central scattering region (including the metallic Cu-MoS$_2$ buffer layer and the semiconductoring BL-MoS$_2$) at equilibrium to display the band alignment of the Cu-MoS$_2$/BL-MoS$_2$ junction for the AC and ZZ directions, respectively. The dark blue regions, approximately 24 to 44 Å (22 to 45 Å) for the AC (ZZ) direction, are low state density regions, indicate the band gap. In the AC direction, the calculated contact potential difference between the Fermi level and the conduction band bottom is 0.14 eV, signifying the creation of a 0.14 eV Schottky barrier, denoted as Φ, as illustrated in Figure 4(a). Similarly, the barrier height for the ZZ direction is 0.12 eV (see Figure 4(b)). The obtained barrier height of Cu-MoS$_2$/BL-MoS$_2$ junction is lower than that of the MoS$_2$/Al Van der Waals heterostructure, which has a barrier height of 0.3 eV.[21] It is also lower than the barrier of the MoS$_2$ heterojunction obtained through phase engineering, with a height of 0.7 to 0.8 eV.[23] The low barrier facilitates the injection of low energy electrons, which is responsible for the current generated by the small bias. Concurrently, the barrier at the interface does not rise, suggesting that an ohmic contact has been formed between the metal and the semiconductor for both directions. In addition, the local space density of states (LDOS) of the device at equilibrium was calculated at $E = 0$ eV, as illustrated in Figure 4(c) and 4(d). At the interface of the metal and the semiconductor (indicated by the dashed red rectangle), electronic states exist at the Fermi level indicates that the metallization of BL-MoS$_2$ and the elimination of the barrier at the contact interface result in the formation of ohmic contact.

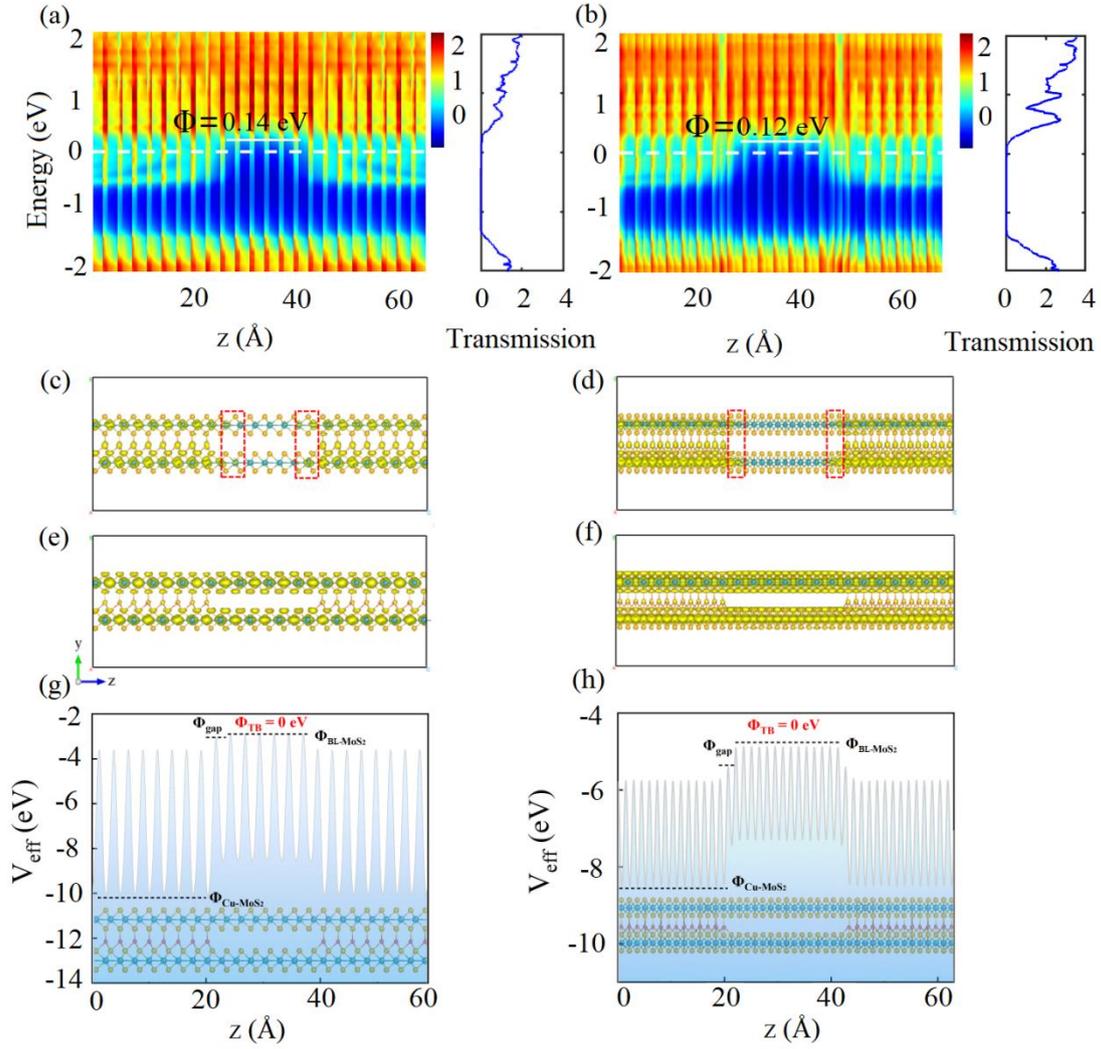

Figure 4. PDOS, transmission, LDOS, and effective potential of the Cu-MoS$_2$/BL-MoS$_2$/Cu-MoS$_2$ devices: PDOS and transmission at equilibrium for (a) the AC direction and (b) the ZZ direction. The values of PDOS are plotted on a logarithmic scale. The white dashed line below indicates the Fermi level, and the white solid line above marks the bottom of the BL-MoS$_2$ conduction band in the central region. The black arrow represents the barrier height. LDOS at $E = 0$ eV along (c) the AC direction and (d) the ZZ direction, as well as the LDOS at $E = 1$ eV along (e) the AC direction and (f) the ZZ direction. Effective potential in (g) AC direction and (h) ZZ direction. $\Phi_{TB}$ represents effective tunnel barrier height, $\Phi_{Cu-MoS_2}$ denotes the minimum V$_{eff}$ that an electron can have in the metal.

The transmission spectra of the equilibrium state are directly affected by the height of the barrier and are shown in the right panels of Figure 4(a) and 4(b). Due to limited electron tunneling, there are no obvious transmission peak in the energy range of -1.36 to 0.12 eV for the ZZ direction and −1.34 to 0.14 eV for the AC direction. At other energies, conduction bands and valence bands of BL-MoS$_2$ can provide channels for electron transport, and corresponding peaks appear in the transmission spectra. When comparing the two directions, it is found that the transmission coefficient in the ZZ direction is higher than that in the AC direction. To clarify the source of this anisotropy, we calculated the LDOS at an energy level of $E = 1$ eV, which clearly demonstrates that there are more transport channels along the ZZ direction than the AC direction, as shown in Figure 4(e) and 4(f), resulting in a larger transmission coefficient for the former.

The tunnel barrier affects the efficiency of carrier injection, which can be evaluated by the effective potential ($V_{eff}$), as shown in Figure 4(g) and 4(h). The tunnel barrier height ($\Phi_{TB}$) is the lowest barrier to be overcome by the electrons when they move at Fermi level from the metal to BL-MoS$_2$.[40] Therefore, $\Phi_{TB}$ can be determined as the difference in $V_{eff}$ between the gap ($\Phi_{gap}$) and BL-MoS$_2$ ($\Phi_{BL-MoS_2}$). Notably, the tunneling barrier vanishes in both directions because $\Phi_{BL-MoS_2}$ exceeds $\Phi_{gap}$. High carrier injection efficiency is expected for the Cu-MoS$_2$/BL-MoS$_2$ contact, which indicates the achievement of low contact resistance. Therefore, low Schottky barrier, the absence of a tunneling barrier and ohmic contact are the reasons for obtaining low contact resistance. Moreover, the obtained contact resistance is below (close to) the theoretical limit of 30 Ω·μm for single-layer materials. The reason lies in our adoption of bilayer MoS$_2$ as the channel material. With a thickness twice that of a single layer, it provides more transport channels.

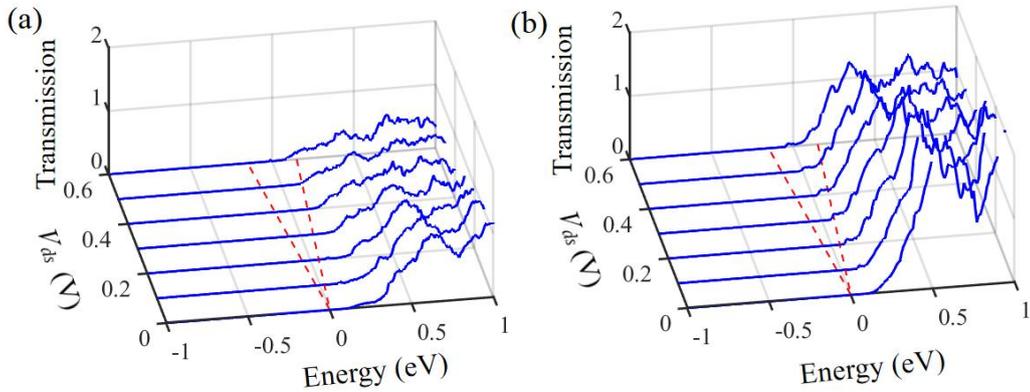

Figure 5. Transmission coefficients versus energy $E$ at different source-drain voltages in (a) AC and (b) ZZ. The dashed red lines represent bias windows.

The characteristics of the *I-V* curves can be more directly explained by the transmission coefficient under different $V_{ds}$, as shown in Figure 5(a) and 2(b). Applying bias voltages causes the source and drain electrode chemical potential ($\mu_L$ and $\mu_R$) to move in opposite energy directions, where the energy range between $\mu_L$ and $\mu_R$ is referred to as the bias window. According to the Landauer-Büttiker formalism, the magnitude of the current is proportional to the integral value of the transmission coefficient in the bias window. When the $V_{ds}$ is small (< 0.3 V), there is no obvious transmission peak in the bias window, and the current is at a minimum. With the increase of $V_{ds}$, the transmission coefficient moves to the low energy region. When the $V_{ds}$ is greater than 0.3 V for the ZZ direction and 0.4 V for the AC direction, the transmission peak corresponding to the high energy contributed by the bottom of the conduction band enters the bias window, which results in a significant increase in current. When the $V_{ds}$ is the same, the transmission peak in the ZZ direction is always greater than the AC direction. As a result, the ZZ direction's current is always larger than the AC direction's current.

To clarify the gate regulation of electrical properties in the 2D semiconductor devices, a vertical electric field was exerted on the BL-MoS$_2$ featuring a HfO$_2$ dielectric layer, as shown in Figure 2. The transfer characteristics curves of Cu-MoS$_2$/BL-MoS$_2$/Cu-MoS$_2$ devices with a fixed $V_{ds}$ of 0.3 V are shown in Figure 6. The current increases with increased positive gate voltages ($V_{gs}$) and decreases with increased negative gate voltage, exhibiting typical n-type conducting behavior. For ZZ direction, the $I_{on}$ can satisfy the ITRS HP standard requirement (900 $\mu$A/$\mu$m),[41] reaching 1176 $\mu$A/$\mu$m before any gate voltage is applied. Under the modulation of -4.5V gate voltage, the device can quickly reach the current of 0.1 $\mu$A/$\mu$m, that is the OFF state current of ITRS HP standard, specified by the ITRS for the high-performance device applications projected for the year 2028.

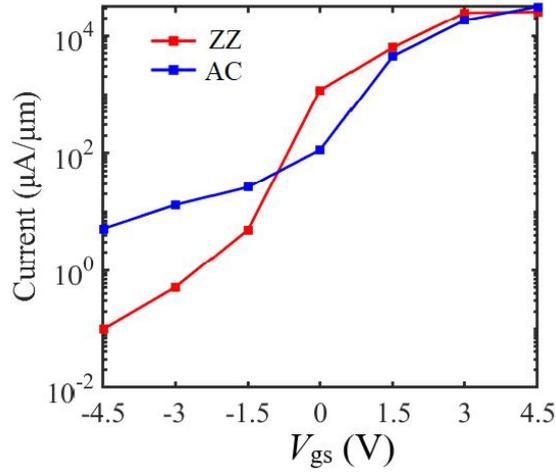

Figure 6. Transfer characteristics of Cu-MoS$_2$/BL-MoS$_2$/Cu-MoS$_2$ devices along the AC and ZZ directions with $V_{ds}$ = 0.3 V.

In order to reveal the origins of the gate regulation performance, we plot the PDOS and the transmission coefficients of the AC and ZZ directions under $V_{gs}$ = -1.5 V, 0 V, and 1.5 V. As shown in the left panels of Figure 7(b) and (e), the conduction band bottom is located in the bias window, which provides a channel for electron transport at $V_{gs}$ = 0 V. In the transmission spectra, the transmission peak contributed by the conduction band bottom appears in the bias window, leading to a current of a certain magnitude with 113 $\mu A/\mu m$ and 1176 $\mu A/\mu m$ for AC and ZZ directions, respectively. When $V_{gs}$ is applied, the electrostatic potential caused by the $V_{gs}$ shifts the energy level in the channel region, causing a gradual change in the conduction band minimum in the central region of BL-MoS$_2$. At $V_{gs}$ = 1.5 V, more of the bottom of the conduction band enters the bias window, providing more channels and leading to an increase in the transmission peaks within the bias window. This, in turn, results in an increase in current (see Figure 7(c) and (f)). When $V_{gs}$ is set to -1.5 V, a higher Schottky barrier is formed, as illustrated in Figure 7(a) and (d). This leads to a reduction in transmission and, consequently, a decrease in current. It is worth noting that the transmission for AC is greater than for ZZ. This is because the electron tunneling behavior through the Schottky barrier near the Fermi level contributes more significantly to the current in the AC direction (see the inset in Figure 7(a) and (d)). This is likely due to the AC channel length being slightly shorter than that of ZZ, resulting in a larger tunneling current for AC. This increase in tunneling current may disrupt the gate's control over the channel, leading to a decrease in device performance.

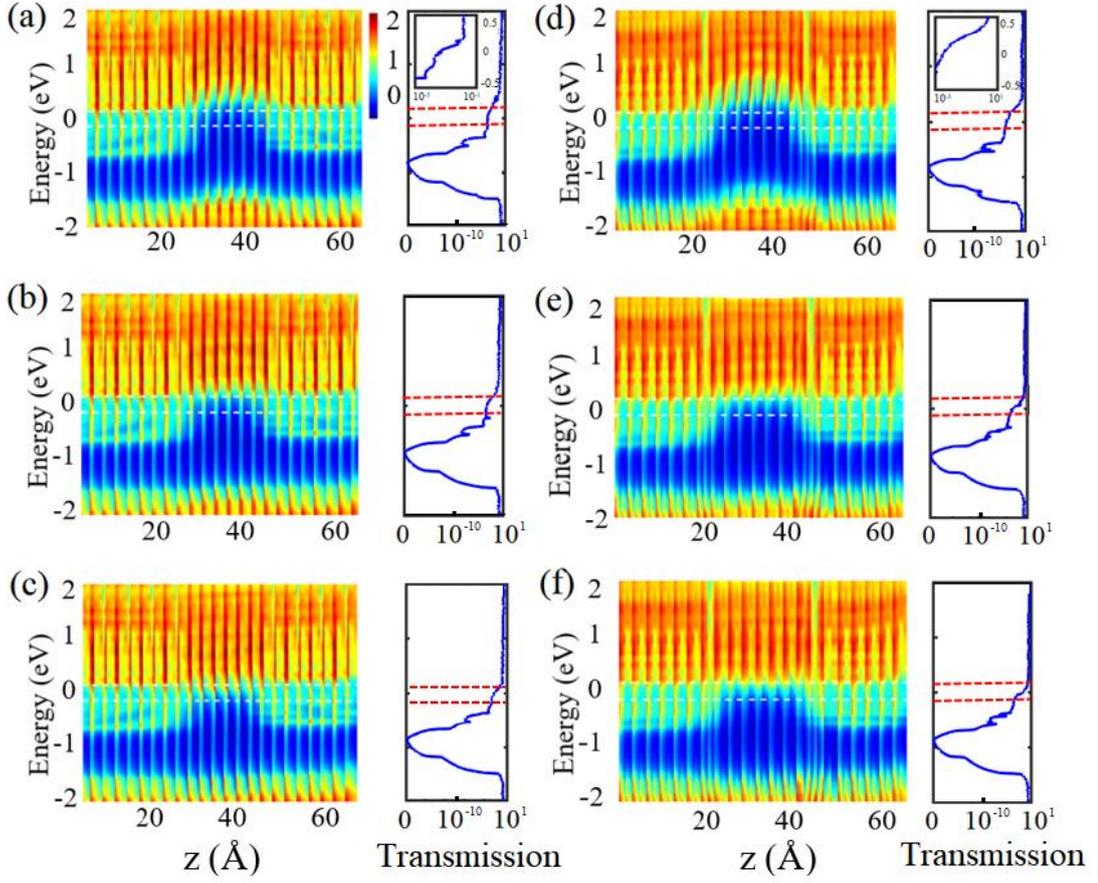

Figure 7. PDOS and Transmission coefficients on log $y$ scale at different gate voltages with $V_{ds} =$ 0.3 V. (a) $V_{gs} = -1.5$ V, (b) $V_{gs} = 0$ V and (c) $V_{gs} = 1.5$ V for the AC direction; (d) $V_{gs} = -1.5$ V, (e) $V_{gs} = 0$ V and (f) $V_{gs} = 1.5$ V for the ZZ direction. The white dashed lines and red dashed lines denotes the bias window. Inset: The transmission coefficients from $E = -0.5$ V to $E = 0.5$ V.

## IV. CONCLUSIONS

In summary, we proposed 2D MoS$_2$-based FETs in which the semiconducting BL-MoS$_2$ served as the central scattering region and metallic Cu-MoS$_2$ as the electrodes. The transport properties of Cu-MoS$_2$/BL-MoS$_2$/Cu-MoS$_2$ were studied by DFT-NEGF method. When the $V_{ds}$ is 0.6 V, the obtained contact resistance is 16.7 Ω·μm in the ZZ direction and 30 Ω·μm in the AC direction. The low contact resistance arises mainly from the low Schottky barrier, the absence of a tunneling barrier, and the formation of an ohmic contact between Cu-MoS$_2$ and BL-MoS$_2$. The regulating effect of gate voltage is significant. The on-state and off-state currents in the ZZ direction are 1176 μA/μm and 0.1 μA/μm, respectively, meeting the current requirements of ITRS HP. Intercalation design addresses the contact issue arising from the Van der Waals gap between metal and 2D semiconductors, thereby solving the prevalent problem of high contact resistance in 2D material devices and fulfilling the ITRS 2028 contact resistance requirements.

## Acknowledgments

This research was supported by the Foundation of Heilongjiang Province Natural Science (grant no. PL2024A003), and the Youth Innovative Talent Project of the Basic Scientific Research Operating Expenses of Provincial Undergraduate Universities in Heilongjiang Province (grant no. 145209215). Haitao Yin thanks Liu Niu for VASP calculaiton.